\documentclass[conference]{IEEEtran}
\IEEEoverridecommandlockouts
\usepackage{cite}
\usepackage{amsmath,amssymb,amsfonts, bm}
\usepackage{algorithmic}
\usepackage{graphicx}
\usepackage{booktabs}
\usepackage{textcomp}
\usepackage{amsmath,graphicx}
\usepackage[dvipsnames]{xcolor}
\usepackage{tabularx}
\usepackage[caption=false,font=footnotesize]{subfig}
\definecolor{darkred}{RGB}{120, 8, 0}

\usepackage[english]{babel}

\graphicspath{ {./Figures/} }
\def\BibTeX{{\rm B\kern-.05em{\sc i\kern-.025em b}\kern-.08em
    T\kern-.1667em\lower.7ex\hbox{E}\kern-.125emX}}

\title{
Bayesian Nonparametric Dimensionality Reduction of Categorical Data for Predicting Severity of COVID-19 in Pregnant Women}

\makeatletter
\newcommand{\linebreakand}{%
\end{@IEEEauthorhalign}
\hfill\mbox{}\par
\mbox{}\hfill\begin{@IEEEauthorhalign}
}
\makeatother

\author{Marzieh Ajirak,$^{\rm \diamond}$ Cassandra Heiselman,$^{\rm \ast}$ Anna Fuchs,$^{\rm \ast}$ Mia  Heiligenstein,$^{\rm \ast}$ Kimberly Herrera,$^{\rm \ast}$\\ Diana Garretto,$^{\rm \ast}$ and Petar M. Djuri\'{c},$^{\rm \diamond}$
\\$^{\rm \diamond}$Department of Electrical and Computer Engineering, Stony Brook University \\
$^{\rm \ast}$Department of Obstetrics, Gynecology and Reproductive Medicine\\
Stony Brook, NY 11794, USA
\thanks{The authors thank the support of NIH under Award RO1HD097188-01.}
}

\begin{document}

\maketitle

\begin{abstract}
The coronavirus disease  (COVID-19) has rapidly spread throughout the world and while pregnant women present the same adverse outcome rates, they are underrepresented in clinical research. We collected clinical data of 155 test-positive COVID-19 pregnant women at Stony Brook University Hospital. Many of these collected data are of multivariate categorical type, where the number of possible outcomes grows exponentially as the dimension of data increases. We modeled the data within the unsupervised Bayesian framework and mapped them into a lower dimensional space using latent Gaussian processes. The latent features in the lower dimensional space were further used for predicting if a pregnant woman would be admitted to a hospital due to COVID-19 or would remain with mild symptoms. We compared the prediction accuracy with the dummy/one-hot encoding of categorical data and found that the  latent Gaussian process had better accuracy. 
\end{abstract}


\section{Introduction}
The coronavirus disease 2019 (COVID-19) has become an unprecedented public health crisis. 
Around the world, many governments issued a call to researchers in machine learning (ML) and artificial intelligence (AI) to address high-priority questions related to COVID-19. 
This call was not unusual because ML methods are finding many uses in medical diagnosis applications. The ML field is rich with examples where  based on predictive models one can estimate disease severity \cite{jiang2020towards} and consequently, the state of a patient's health \cite{yao2020severity, alimadadi2020artificial, miotto2016deep}. These models employ data-driven algorithms that can extract features and discover complicated patterns that could have not been recognized or interpreted by humans.

Pregnant women are a particularly important patient population to study due to their vulnerability to disease and the often underrepresentation of the population in clinical research  \cite{Favre2020}. Despite studies in this field \cite{breslin2020covid, wu2020perinatal, chen2020clinical, dashraath2020coronavirus, pierce2020clinical}, there has been a relative sparsity of data in regards to COVID-19 and its effect on pregnancy. Utilizing ML techniques to study this population during the pandemic can help build pregnancy-specific evidence to guide clinical recommendations \cite{chen2020early}.

Much of the medical data are of multivariate categorical type, and typically they represent patients' demographics, maternal comorbidities, pregnancy complications, and disease symptoms. As a result, one has to work with long vectors of categorical variables which in turn leads to a huge number of possible realizations. This then creates very sparse spaces when we deal with a limited number of data \cite{agresti2018introduction}.
Besides, the data include random errors and systematic biases, and sometimes  they are missing \cite{little2019statistical}. 
By overcoming the challenges that clinical data introduce, one can layout the grounds for developing more accurate models and efficient algorithms for inference.

The key to have successful predictive methods largely depends on feature selection and data representation. A common approach is to have a clinical doctor specify the variables and label the clinical data to be used as training sets. Then the ML method will find mappings and features from the data, which subsequently will be tested on new data sets. Although appropriate in many situations, a supervised definition of the features contributes to losing an opportunity to learn latent patterns and features \cite{miotto2016deep}. In countering the subjectivity of defining the features, an unsupervised learning approach can be used to extract useful information from data. One other advantage of unsupervised learning is that abstract features of patients can often be represented in low-dimensional spaces and thus, they can summarize efficiently the information available in the data. This further allows for easy visualization of the cohort of patients under consideration.

In the ML literature, categorical latent Gaussian processes provide data efficient and powerful Bayesian framework for learning latent functions or patterns \cite{gal2015latent}. In this paper, we model the categorical data from pregnant women as generated non-linearly from a latent space. More specifically, we map the categorical variables including maternal comorbidities, pregnancy complications, ABO blood types, etc., into a continuous lower dimensional space. Then we use these learned features along with the remaining numerical data (maternal age, BMI, etc.) to predict whether (a) the patient will develop severe symptoms and will come back to the hospital due to COVID-19, days after tested positive, or (b) the patient will remain asymptomatic or symptomatic but with mild symptoms.
We compared the performance obtained by direct and non-linear dimensionality reduction of the categorical data with the methodology of one-hot encoding, which is  commonly applied in the machine learning circles when dealing with categorical data. 

The remainder of this paper is organized as follows. In the next section, we explain the categorical latent Gaussian model first introduced in \cite{gal2015latent}. Then we introduce an alternative pipeline that deals with categorical data. We test the proposed approach first on synthesized data and then on original COVID-19 data. 

\begin{table}[]
    \caption{List of Symbols and Notations}
    \label{tab:my_label}
    \centering
    \begin{tabular}{r c p{0.3\textwidth}}
    \toprule
    $N$ & -- & Number of patients\\
    $D$ & -- & Number of categorical variables \\
    $K$ & -- & Number of possible outcomes\\
    $Q$ & -- & Dimensionality of the latent space\\
    $\bm{x}_n$   &  -- & Latent value of the $n$th patient\\
    ${\bm y}_n$&-- &A $D-$dimensional observed vector \\
    $y_{nd}$&--& The $d$th element of ${\bm y}_n$ that can take one of $D$ categorical values \\ 
    $\bm{f}_{nd}$  &  -- & Vector of probability weights of the $d$th categorical variable for the $n$th patient\\
    $\mathbf{K}_d$ & -- & Covariance matrix corresponding to the $d$th categorical variables\\
    $\kappa$& -- & A kernel function\\
    $\mathcal{F}$ & --&A sample function from $\mathcal{GP}(., .)$ \\
    $\bm{Z}$ & -- & Inducing input locations\\
    $\bm{U}$ & -- & Inducing variables\\
    \bottomrule
    \end{tabular}
\end{table}

\section{Background on GPLVM}
Gaussian process latent variable models (GPLVMs) are Bayesian nonparametric frameworks that allow for unsupervised learning \cite{titsias2010bayesian}. GPLVMs can be seen as multi-output Gaussian process regressions when the inputs are unobserved. To be more specific, let $\bm{f}_{n} \in \mathbb{R}^{K}$ be the $n$-th observed data vector of dimension $K$, $n=1, 2, \ldots, N$. Further, let these data be associated with inputs $\bm{x}_n \in \mathbb{R}^{Q}$ through $K$ different functions. 
If we assume that these functions are independent, then for  $\bm{f}_n$ we can write
\begin{align}
p \left(\bm{f}_{n} (\bm{x}_{n})\right)=\prod_{k=1}^{K} p\left(f_{nk} (\bm{x}_{n})\right),
\end{align}
where $f_{nk}(\bm{x}_{n})$ represents the $k$th dimension of $\bm{f}_{n}(\bm{x}_{n})$ and 
\begin{align}
p\left(f_{nk}(\bm{x}_{n})\right)=\mathcal{N}\left(f_{nk} ; 0, {k}({\bm x}_n,{\bm x}_n^\prime)\right),
\end{align}
where the notation ${\cal N}(f_{nk}; 0, {k}({\bm x}_n,{\bm x}_n^\prime))$ means that the random variable $f_{nk}(\bm{x}_{n})$ is Gaussian with mean zero and variance defined by the covariance function ${k}({\bm x}_n,{\bm x}_n^\prime)$. 
In order to automatically learn the dimensionality of the latent space, we will use the concept known as Automatic Relevance Determination (ARD) with the kernel 
\begin{align}
\kappa\left(\bm{x}_n, \bm{x}_n^{\prime}\right)=\sigma_{f}^{2} \exp \left(-\frac{1}{2} \sum_{q=1}^{Q} \alpha_{q}\left(x_{n,q}-x_{n,q}^{\prime}\right)^{2}\right).
\end{align}
In GPLVMs, ${\bm X}\in {\mathbb R}^{N\times Q}$ is a matrix of latent variables, and therefore we assign it a prior density. A typical approach is to use the standard Gaussian distribution, and thus we have 
\begin{equation}
\label{eq: lvm x}
  p(\bm{X})=\prod_{n=1}^{N} \mathcal{N}\left(\bm{x}_{n} ; \bm{0}, \bm{I}_{Q}\right),
\end{equation}
where the $\bm{x}_n$'s are the rows of ${\bm X}$. By defining the matrix of observations $\bm{F} \in \mathbb{R} ^ {N \times K}$, where the rows represent  the multiple outputs ${\bm f}_n$, we wish to compute the marginal likelihood of the data:
\begin{equation}
p(\bm{F})=\int p(\bm{F} | \bm{X}) p(\bm{X}) {\rm d} {\bm X}.
\end{equation}

The authors in \cite{titsias2010bayesian} developed a variational Bayesian approach for the marginalization of the latent variables, $\bm X$, allowing them to optimize the resulting lower bound on the marginal likelihood with respect to the hyperparameters. They further used the  lower bound for model comparison and automatic selection of the latent dimensionality.

\section{Multivariate Discrete GPLVM}
\subsection{Generative Model}
We consider now the discrete version of GPLVM where for each input $\bm{x}_{n}$,  we observe a discrete variable $y_{n}$ that can take values $1, ..., K$, with probabilities
\begin{equation}
\label{eq:prob}
p\left(y_{n}=k\right)=\frac{\exp \left(f_{nk}\right)}{\sum_{k^{\prime}=1}^{K} \exp \left(f_{ nk^{\prime}}\right)}.
\end{equation}

\begin{figure}[!htbp]
    \centering
    \includegraphics[width=.48\textwidth]{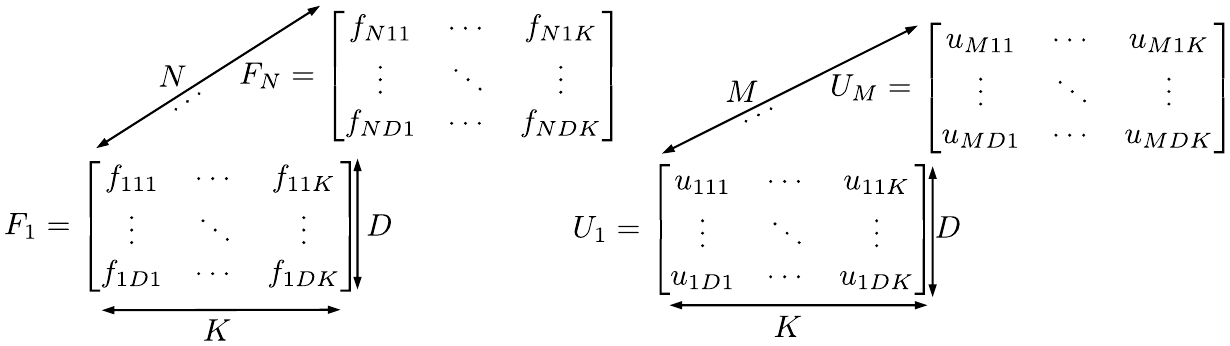}
    \caption{Latent weights and inducing variables.}
    \label{fig: FU}
\end{figure}

\begin{figure}[!htbp]
    \centering
    \includegraphics[scale=0.6]{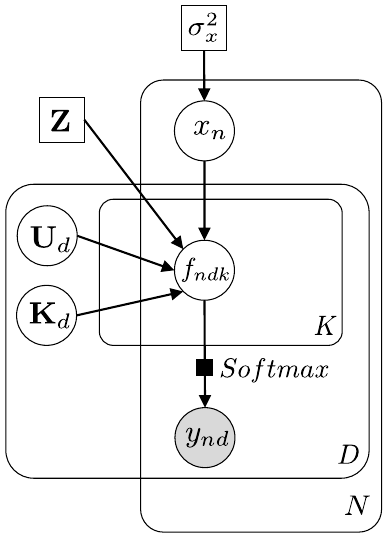}
    \caption{A graphical representation of the generative model.}
    \label{graph1}
\end{figure}

In the multivariate case, we have $\bm{y}_n \in \mathbb{R}^{D}$. Next, we consider a generative model for a dataset $\bm{Y} \in \mathbb{R}^{N\times D}$ with $N$ observations and $D$ categorical variables. We denote the $d$-th  variable in the $n$-th observation by $y_{n d}.$ Now we express \eqref{eq:prob} as 
\begin{equation}
p\left(y_{nd}=k\right)=\frac{\exp \left(f_{ndk}\right)}{\sum_{k^{\prime}=1}^{K} \exp \left(f_{ ndk^{\prime}}\right)},
\end{equation}
where $f_{n d k}$ is function of the input variable $\bm{x}_{n} \in \mathbb{R}^{Q}, $ i.e.,
$f_{ndk}= \mathcal{F}_{d k}\left(\bm{x}_{n}\right).$  
Next, we summarize the generative model (the indices below have the following meaning:  $n$ refers to observation, $d$ to the dimension of the output, $m$ to an inducing point (defined below), and $k$ to category),  
\begin{align}
\label{eq:x}
x_{nq} &\stackrel{\text { iid }}{\sim} \mathcal{N}\left(0, \sigma^{2}_{x}\right),\\
\label{eq:F}
\mathcal{F}_{d k} &\stackrel{\text { iid }}{\sim} \mathcal{GP}\left(0, {k}_{d}(\cdot, \cdot)\right),\\
f_{n d k}&=\mathcal{F}_{d k}\left(\bm{x}_{n}\right),\\
u_{m d k}&=\mathcal{F}_{d k}\left(\bm{z}_{m}\right),\\
\label{eq:softmax}
p(y_{nd}=k) &= \frac{\exp \left(f_{ndk} \right)}{\sum_{k^{\prime}=1}^{K} \exp \left(f_{ndk^{\prime}}  \right)},
\end{align}
where ${x}_{nq}$ and $\mathcal{F}_{dk}$ are latent variables with prior distributions given by \eqref{eq:x} and \eqref{eq:F}, respectively, with ${\cal GP}$ signifying Gaussian process (GP). Further, the $\bm{z}_m$s are inducing inputs, $m=1, 2 ,\ldots, M$, and the $u_{mdk}$s are inducing outputs whose role is explained further below. We note that we assume a Gaussian distribution prior with standard deviation $\sigma_{x}^{2}$ for ${x}_{nk}$, and a GP  prior for each of the functions $\mathcal{F}$. 
We reiterate that for each vector of latent function values $\bm{f}_{dk}$,  we introduce a separate set of $M$ variational inducing variables $\bm{u}_{dk}$, evaluated at a set of inducing input locations from the set ${\cal Z}=\{\bm{z}_1, \bm{z}_2, \ldots, \bm{z}_M\}$. It is assumed that all $\bm{u}_{dk}$s are computed at the same inducing locations. The inducing variables are  function points drawn from the GP prior and lie in the same latent space as  $F$ variables (Fig. \ref{fig: FU}). The pictorial description of the generative model is displayed in Fig. \ref{graph1}.

\subsection{Inference}
The marginal log-likelihood is intractable because of the covariance function of the GP and the nonlinear Softmax likelihood in \eqref{eq:softmax}.
We consider a variational approximation to the posterior distribution of $\bm{X}, \bm{F}$ and $\bm{U}\in{\mathbb R}^{M\times D\times K}$ factorized as,
\begin{equation}
q(\bm{X}, \bm{F}, \bm{U})=q(\bm{X}) q(\bm{U}) p(\bm{F} | \bm{X}, \bm{U}).
\end{equation}
By applying Jensen’s inequality, we can write a lower bound of the log-evidence (ELBO) as
\begin{multline}
\label{elbo}
 \log p(\bm{Y})=
\log \int p(\bm{X}) p(\bm{U}) p(\bm{F} | \bm{X}, \bm{U}) p(\bm{Y} | \bm{F}) \mathrm{d} \bm{X} \mathrm{d} \bm{F} \mathrm{d} \bm{U} \\
\geq - \mathrm{KL}(q(\bm{X}) \| p(\bm{X}) )-\mathrm{KL}(q(\bm{U}) \| p(\bm{U})) \\
\quad\quad +\sum_{n=1}^{N} \sum_{d=1}^{D} \int q\left(\bm{x}_{n}\right) q\left(\bm{U}_{d}\right) p\left(\bm{f}_{n d} | \bm{x}_{n}, \bm{U}_{d}\right)\\
\cdot \log p\left(\bm{y}_{n d} | \bm{f}_{n d}\right) \mathrm{d} \bm{x}_{n} \mathrm{d} \bm{f}_{n d} \bm{U}_{d}
:=\mathcal{L},
\end{multline}
where,
\begin{multline}
\label{fnd}
p\left(\bm{f}_{n d} | \bm{x}_{n}, \bm{U}_{d}\right)=
\prod_{k=1}^{K} \mathcal{N}(f_{ndk};\mathbf{k}_{d, n M}^\top \mathbf{K}_{d, M M}^{-1}  \bm{u}_{d k},\\
{k}_{d, n n}-\mathbf{k}_{d, n M}^\top \mathbf{K}_{d, M M}^{-1} \mathbf{k}_{d, M n}).
\end{multline}

The lower bound is still intractable because of the softmax likelihood, $\log p\left(\bm{y}_{n d} \mid \bm{f}_{n d}\right)$.
Therefore, we will compute the lower bound $\mathcal{L}$ and its derivatives with the Monte Carlo method. We draw samples of $\bm{x}_{n}, \bm{U}_{d}\in {\mathbb R}^{M\times K}$ (see Fig. \ref{fig: FU}) and $\bm{f}_{n d}$ from $q\left(\bm{x}_{n}\right), q\left(\bm{U}_{d}\right),$ and $p\left(\bm{f}_{n d} \mid \bm{x}_{n}, \bm{U}_{d}\right)$, respectively, and estimate $\mathcal{L}$ with the sample average.
We consider mean field variational approximation of the latent points $q(\bm{X})$ and a joint Gaussian distribution for $q(\bm{U})$ as,
\begin{equation}
q(\bm{U})=\prod_{d=1}^{D} \prod_{k=1}^{K} \mathcal{N}\left(\bm{u}_{ d k} ; \boldsymbol{\mu}_{d k}, \bm{\Sigma}_{d}\right),
\end{equation}
\begin{equation}
q(\bm{X})=\prod_{n=1}^{N} \prod_{q=1}^{Q} \mathcal{N}\left(x_{n q} ; m_{n q}, \sigma_{n q}^{2}\right),
\end{equation}
where the covariance matrix $\bm{\Sigma}_{d}$ is shared for the same categorical
variable $d$. The KL divergence in $\mathcal{L}$ can be computed analytically with the given variational distributions.
We need to optimize the hyperparameters of each GP (parameters of $\mathbf{K}_d$), parameters of the variational random variables $\bm{u}_{dk}$, $\boldsymbol{\mu}_{dk}$, $\bm{\Sigma}_d$, mean $m_{nq}$ and variance $\sigma^2_{nq}$ of the latent inputs.

\section{Experiments and results}
\subsection{1-D Input and Output}
Consider the categorical variable $y$ that can take values 001, 010, and 100 (or blue, red, and green). The input variable $x$ comes from a space of patients. Further, let three functions $\mathcal{F}_{11}, \mathcal{F}_{12}$, and $\mathcal{F}_{13}$ model the $f_{ndk}$s  that are used for computing the probability of each category (the first index of the functions refers to the dimension, which in this example equals to one). For instance, $\mathcal{F}_{11}(x_n)$ is proportional to the probability of $y$ for patient $n$ with input $x_n$ being 001.
Similarly, we define ${f}_{n12}$ and ${f}_{n13}$ for the categories 010 and 100, respectively.

\begin{figure}[!hbt]
    \centering
    \subfloat[Embedding the categorical variables into lower dimension space. ]{\includegraphics[width=0.48\textwidth]{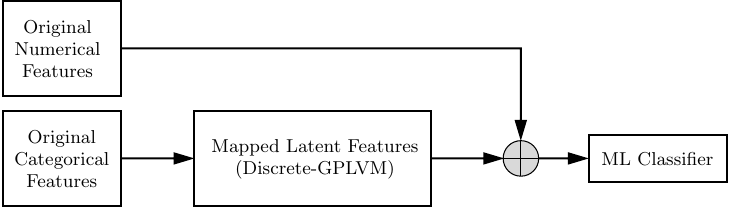}}\\
    \subfloat[One-hot encoding of the categorical variables before dimensionality reduction. ]{\includegraphics[width=0.48\textwidth]{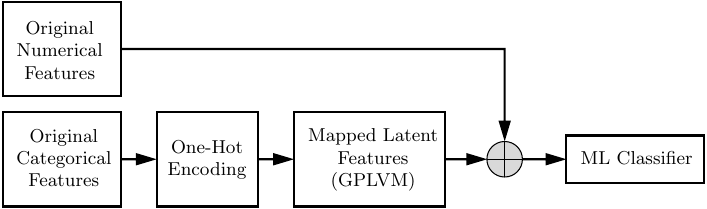}}
    \caption{Model Pipeline}
    \label{pipeline}
\end{figure}

\begin{figure}[!hbt]
    \centering
    \subfloat[Generative model for a single categorical variable $y$.]{
    \includegraphics[width=0.42\textwidth]{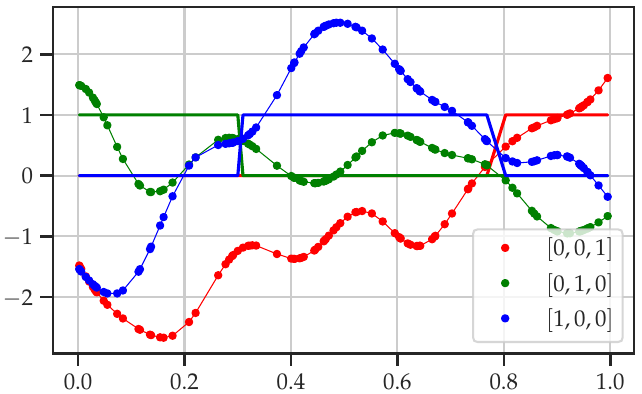}}\\
    \subfloat[Learned latent variables with Discrete-GPLVM]{\includegraphics[width=0.45\textwidth]{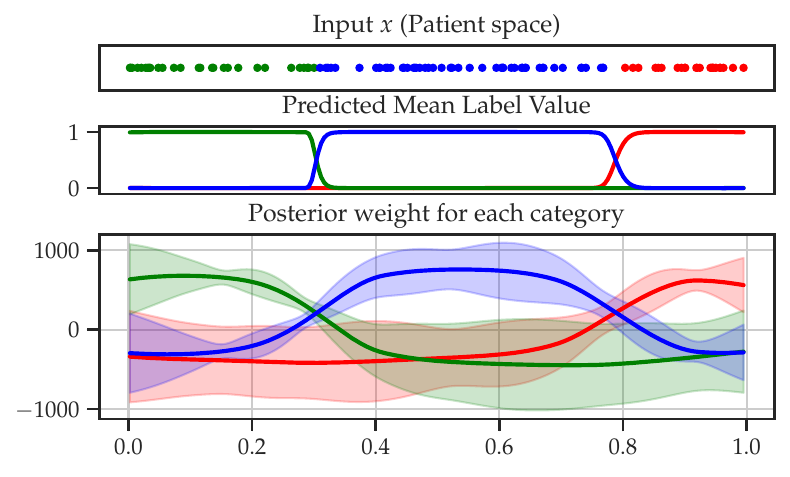}}\\
    \subfloat[Learned latent space of $x$ with One-Hot Encoding-GPLVM.]{\includegraphics[width=0.4\textwidth]{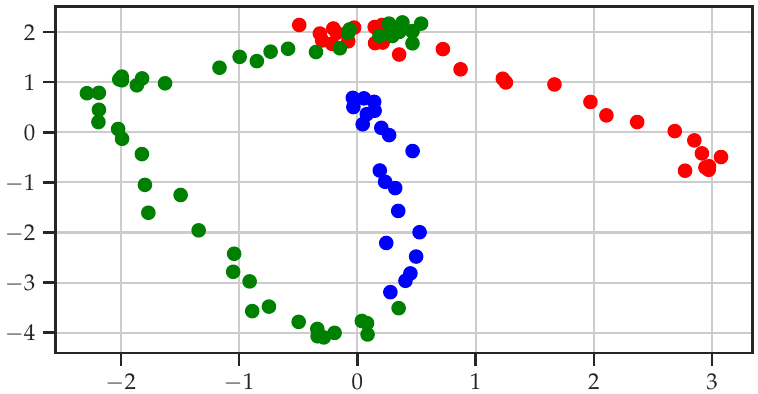}}
    \caption{Synthetic Example}
    \label{fig:classification}
\end{figure}

We perform the inference using the introduced method (Fig. \ref{pipeline} (a)) and compare it with the one-hot encoding of the categorical variables (Fig. \ref{pipeline} (b)).
 We observe that by the one-hot encoding and then applying GPLVM, the structure of the latent space is distorted. The first two dimensions of $x$ are shown in Fig. \ref{fig:classification} (c). Although a  one-dimensional manifold is detected, the points at the boundary of the two clusters are obviously distorted.

\label{sec:experiment}

\subsection{COVID-19 Data}

\begin{table}[!t]
\centering
\caption{List of COVID-19 patients variables}
\begin{tabular}{l r}
\toprule
\textbf{Input Variables}& \textbf{Type\quad\quad}\\
\midrule
Maternal age (year) & Numerical\\ 
BMI (kg/m2) & Numerical\\
Gravidity & Numerical \\
Parity & Numerical \\
Admission Lab Values & Numerical \\
\midrule
Thermodynamic Symptoms  & Categorical\\
Lower Respiratory Symptoms  & Categorical \\
HEENT Symptoms & Categorical\\
GI symptoms & Categorical\\
Hemodynamic Symptoms & Categorical\\
Cardiovascular Symptoms & Categorical\\
Musculoskeletal Symptoms & Categorical \\
Race &  Categorical\\
Employer type & Categorical\\
Insurance & Categorical\\
Known sick contact type & Categorical\\
Maternal comorbidities & Categorical\\ 
Pregnancy complications  & Categorical\\
ABO blood type & Categorical \\
\midrule
Symptoms at time of diagnosis & Binary (Yes/No)\\
Admitted to hospital for COVID-19 & Binary (Yes/No)\\
Admitted to ICU & Binary (Yes/No)\\
\bottomrule
\end{tabular}

\label{tab: variables}
\end{table}

We used data collected at SBUH of 155 test-positive COVID-19 pregnant women. The dataset is composed of  categorical variables including patients’ symptoms, maternal comorbidities, pregnancy complications, race, employer type, insurance, known sick contact, and ABO blood type. It also has  numerical data including age, BMI, gravidity, parity, and admission lab values. The list of categorical and numerical variables is summarized in Table \ref{tab: variables}. The  cohort  consisted  of  60  asymptomatic  cases,  81  moderate  symptomatic, and  14 severely  ill  patients  who  were  admitted  to  hospital  for COVID-19. Of the latter 14 patients, four were admitted to ICU. 

We first reduced the dimension of categorical data by mapping them into a lower-dimension space using discrete-GPLVM. Next, we used the extracted latent features combined with numerical variables for the supervised task of binary classification. Then we converted the categorical variables to one-hot features and then applied GPLVM. For classification we employed Random Forest, Na\"{\i}ve Bayes, AdaBoost, $k$-Nearest Neighbours (kNN), Support Vector Machine (SVM), and Logistic Regression. We compared the performances of the methods by Area Under the ROC Curve (AUC), Classification Accuracy (CA), F1, Precision, and Recall, where
\begin{equation}
 \mathrm{CA}= \frac{TP+TN}{TP+TN+FP+FN},  
\end{equation}
with $TP$ representing True Positive, $TN$ True Negative, $FP$  False Positive, and $FN$ False Negative predictions, 
\begin{equation}
\mathrm{Recall}=\frac{TP}{TP+FN},
\end{equation}
\begin{equation}
\mathrm{Precision}=\frac{TP}{TP+FP},
\end{equation}and
\begin{equation}
\mathrm{F1}=2*\frac{\mathrm{Recall}*\mathrm{Precision}}{\mathrm{Recall}+\mathrm{Precision}}.
\end{equation}
The results are summarized in Tables \ref{tab:results1} and \ref{tab:results2}.
\begin{table}[!ht]
\caption{Discrete-GPLVM}
\label{tab:results1}
\centering
\begin{tabular}{llllll} 
\toprule
Model & AUC & CA & F1 & Precision & Recall \\
\midrule
Random Forest &0.842 & 0.787 & 0.788 & 0.789 & 0.787 \\
Na\"{\i}ve Bayes & 0.781 & 0.639 & 0.631 & 0.751 & 0.639 \\
AdaBoost & 0.715 & 0.723 & 0.724 & 0.728 & 0.723 \\
kNN & 0.674 & 0.671 & 0.655 & 0.661 & 0.671 \\
SVM & 0.670 & 0.600 & 0.604 & 0.613 & 0.600 \\
Logistic Regression & 0.601 & 0.619 & 0.611 & 0.608 & 0.619\\
\bottomrule
\end{tabular}
\end{table}

\begin{table}[!ht]
\caption{One-Hot Encoding-GPLVM}
\label{tab:results2}
\centering
\begin{tabular}{llllll}
\toprule
Model & AUC & CA & F1 & Precision & Recall \\
\midrule
Random Forest & 0.770 & 0.729 & 0.729 & 0.729 & 0.729 \\
SVM & 0.702 & 0.677 & 0.642 & 0.678 & 0.677 \\
Na\"{\i}ve Bayes & 0.689 & 0.568 & 0.550 & 0.692 & 0.568 \\
kNN & 0.677 & 0.665 & 0.647 & 0.653 & 0.665 \\
AdaBoost & 0.671 & 0.684 & 0.685 & 0.687 & 0.684 \\
Logistic Regression & 0.607 & 0.606 & 0.601 & 0.598 & 0.606\\
\bottomrule
\end{tabular}
\end{table}

The results suggests that the performance of almost all classifiers improved by using the discrete GPLVM. The best performance of all classifiers was achieved by Random Forest. It appears that with dimensionality reduction using discrete GPLVM we compress information better than with GPLVM carried out by one-hot encoding.      

\begin{figure}[!htbp]
    \centering
    \includegraphics[width=0.9\linewidth]{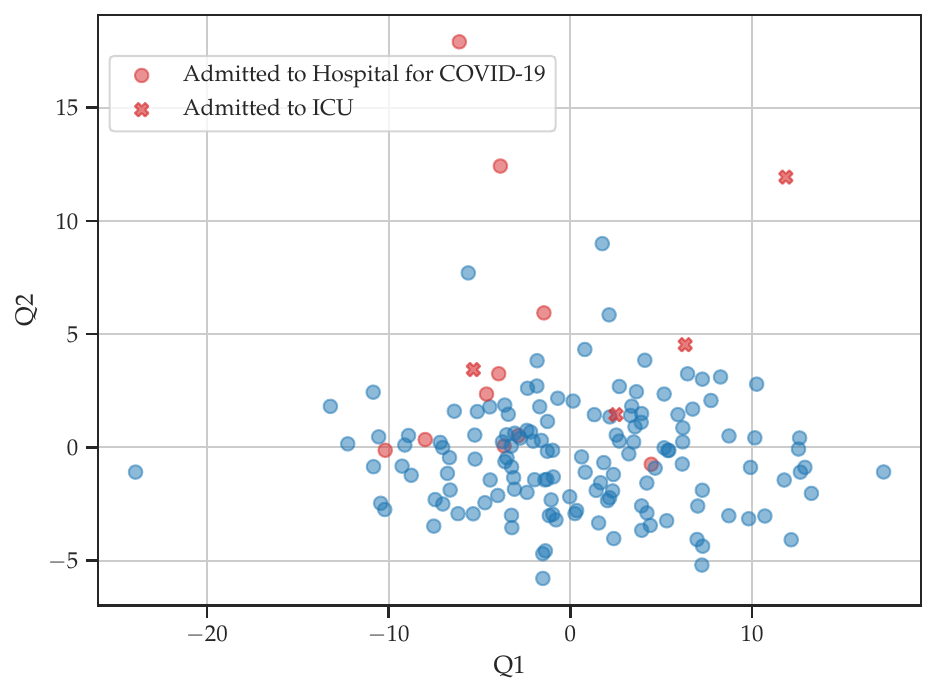}
    \caption{Visualization of the patients. Blue circles represent asymptomatic patients or patients with mild symptoms, red circles represent patients who were hospitalized and red crosses are patients who were admitted to ICU.}
    \label{fig: visualization}
\end{figure}

We also mapped the data for the task of visualization of the cohort. 
Figure \ref{fig: visualization} shows the visualization of the patients using discrete-GPLVM by setting the latent dimension to $Q=2$. 
We observe that the latent features of the symptomatic patients or patients with mild symptoms (blue circles) are well clustered and somewhat separated from the patients who were hospitalized or who were admitted to ICU (red circles and red crosses).

\section{Conclusion}
\label{sec: conclusion}
In this paper, we modeled multivariate categorical data using Gaussian process latent variable models to predict if a pregnant women would be admitted to the hospital due to COVID-19. In our approach, we used  a data-efficient Bayesian framework for reducing the dimension of high-dimensional categorical data. Our tests with synthetic data showed that the method is capable of finding latent structures of the data.  Further, the results on test-positive COVID-19 pregnant women suggest that the method discovered latent structures that were useful for further classification of the data. 
\bibliographystyle{IEEEtran}
\bibliography{refs}

\end{document}